\documentclass[prl,twocolumn]{revtex4-1}
\usepackage[utf8x]{inputenc}
\usepackage{amsbsy}
\usepackage{graphicx}
\usepackage{color}
\usepackage{dcolumn}
\usepackage{amsmath}
\usepackage{amssymb}
\usepackage{amsfonts}
\usepackage{subfigure}
\usepackage{bm}

\usepackage[colorlinks=true, citecolor=black, linkcolor=black, urlcolor=black]{hyperref}

\begin{document}
\title{Universal enhancement of superconductivity in
  two dimensional semiconductors at low doping by electron-electron interaction.}

\author{Matteo Calandra}
\email[]{matteo.calandra@upmc.fr}
\author{Paolo Zoccante}
\author{Francesco Mauri}
\email[]{francesco.mauri@upmc.fr}

\affiliation{IMPMC, UMR CNRS 7590, Sorbonne Universit\'es - UPMC Univ. Paris 06,  MNHN, IRD, 4 Place Jussieu, F-75005 Paris, France}

\begin{abstract}
In 2-dimensional multivalley
semiconductors, at low doping, even a moderate electron-electron interaction
enhances the response to any
perturbation inducing a valley polarization.
If the valley polarization is due to the electron-phonon
coupling, the electron-electron interaction
results in an enhancement of the superconducting
critical temperature. By performing first principles
calculations beyond DFT, we prove
that this effect
accounts for the unconventional doping-dependence 
of the superconducting transition-temperature (T$_c$) and of the
magnetic susceptibility measured in Li$_x$ZrNCl. 
Finally we discuss what are the conditions for a maximal T$_c$
enahnacement
in weakly doped 2-dimensional semiconductors.
\end{abstract}

\pacs{71.10.Ca, 74.20.pq, 63.20.dk, 63.22.Np }
\maketitle
%
%
The quest for high T$_c$ superconductivity has mainly focused on
strongly correlated materials in proximity of electronic instabilities
like the Mott transition (cuprates~\cite{Bednorz})
or fragile magnetic states (iron pnictides~\cite{Kamihara,Canfield}).
Heavily doped 
three dimensional (3D) covalently bonded semiconductors,
like diamond~\cite{Ekimov},
silicon~\cite{Bustarret} and SiC~\cite{Ren,Kriener} have been considered
as an alternative, that, however, has lead, so far, to fairly low T$_c$
($ < 10$ K).
In 3D, the density of states at the Fermi level slowly grows with doping.
As T$_c$ increases with the density of states~\cite{Bustarretdiamond},
a large number of carriers has to be introduced to achieve a
large T$_c$.
This demanding requirement could be released in 2 dimensional (2D) semiconductors, 
such as transition metal dichalcogenides~\cite{NovoselovPNAS,XiaoHeinz,IwasaChiral,YeIwasaMoS2},
cloronitrides~\cite{Yamanaka, Yamanaka2} or other layered materials with massive Dirac fermions, 
where the doping can be controlled by intercalation ~\cite{Yamanaka, Yamanaka2} or field-effect 
~\cite{Ye,IwasaChiral,YeIwasaMoS2}.
In 2D, the density of states is a constant function of the Fermi energy ($\epsilon_F$) and, in principle, 
T$_c$ is expected to be insensitive on doping.
Surprisingly, measurements on Li$_x$ZrNCl ~\cite{Taguchi, Yamanaka, Yamanaka2}, a 
weakly-doped multivalley 2D semiconductor, 
revealed that T$_c$ not only does not increase with doping but even 
decreases.
Here we show that the e-e interaction is responsible for such a puzzling behavior. 
In particular, in a weakly-doped 2D multivalley semiconductor,
e-e manybody effects enhance the response to any
perturbation inducing a valley polarization.
If the valley polarization is due to the electron-phonon
coupling, the e-e interaction will
lead to a large increase of T$_c$.
We demonstrate that this effect
explains the high T$_c$ in Li$_x$ZrNCl and its
unconventional behavior~\cite{Taguchi} as a function of doping.
Finally, by finding the conditions for a maximal Tc enhancement, we
show how weakly-doped 2D
semiconductors are an alternative route towards high Tc superconductivity.
%
%

The electronic structure of multivalley semiconductors has minima
(maxima) in the conduction (valence) band that are named
valleys. In the low doping limit, the
equivalent $g_v$ valleys are occupied by few electrons or holes and the electronic
structure is  described by the effective mass theory. The resulting
model Hamiltonian
is that of a multicomponent electron gas of mass $m^*$
and density $n$
where the valley index plays the role of a pseudospin.
Since at low doping the Fermi momentum $\kappa_F$,
as measured from the valley bottom, is much
smaller than the valley separation,
the intravalley e-e interaction dominates over the
intervalley one, and (for an isotropic mass tensor)
the manybody Hamiltonian has SU$(2 g_v)$ symmetry in valley and spin
indexes \cite{SI,Ando,Marchi,DasSarma2009}.
In 2D the intravalley Coulomb
interactions is $2\pi/(\epsilon_M q)$, where $q$ is
the exchanged momentum, and $\epsilon_M$ the environmental dielectric constant.
The strength of e-e scatteringis measured by the
parameter $r_s=1/(a_B\sqrt{\pi n})$ where
  $a_B=\epsilon_M \hbar^2/(m^* e^2)$.

The magnetic properties of a doped semiconductor are
 described by the interacting spin susceptibility, $\chi_s$:
\begin{equation}
\chi_s\, =\, \frac{\partial M}{\partial {B_{\rm ext}}}
\end{equation}
where $M$ and $B_{\rm ext}$ are the spin magnetization and 
the external magnetic field.
The non-interacting spin susceptibility is 
doping independent, namely $\chi_{0s}=\mu_S^2N(0)$, where
$\mu_S$ is the electron-spin magnetic moment and $N(0)
=g_v m^*/(\pi\hbar^2)$ the density of states at $\epsilon_F$.
Manybody e-e effects increase $\chi_s$
with respect to its
non-interacting value $\chi_{0s}$ and can lead to a magnetic
state. Indeed the e-e energy is lower in the spin-polarized state,
since electrons with same spin and valley
cannot occupy same spatial position because of the Pauli exclusion principle.
The enhancement $\chi_s/\chi_{0s}$ increases with increasing $r_s$ (as the relative contribution of exchange to the total energy increases)
and it is significant already at moderate correlations, $r_s\approx 1$~\cite{Zhang1,Zhang2,Attaccalite,Marchi,DasSarma2009}.

In a multivalley electron gas, an external perturbation can induce
a valley polarization.
The existence of such a valley polarization in 2D
systems is at the heart of
recent developments in the field of valleytronics~\cite{XiaoHeinz}, the
valley analogue of spintronics.
Any perturbation inducing an asymmetry in the population of the
different valleys can then be seen as an external pseudo magnetic field.
In analogy with the magnetic case,
a valley susceptibility $\chi_v$
is defined as the first derivative of the valley magnetization
($\mu_s$ times the valley-population difference) with respect
to the external pseudo magnetic field.

In the low doping limit, because of the SU$(2g_v)$ valley-spin
symmetry of the model Hamiltonian \cite{Ando,Marchi,DasSarma2009},
the valley susceptibility $\chi_v$ is
equal to the spin susceptibility $\chi_s$.
This equality was experimentally
verified in AlAs quantum wells \cite{Gunawan}
 where the pseudo magnetic field
was generated by a strain deformation.
Similarly to the strain case in AlAs,
an intervalley phonon can also act as a pseudo magnetic field
by inducing a valley splitting and a valley polarization via the
electron-phonon interaction. As a consequence, the
manybody enhancement of the valley susceptibility can result in
an augmentation of the superconducting critical
temperature ($T_c$) at low doping, as we show it happens in Li$_x$ZrNCl.
\begin{figure}[t]
    \includegraphics[width=8cm]{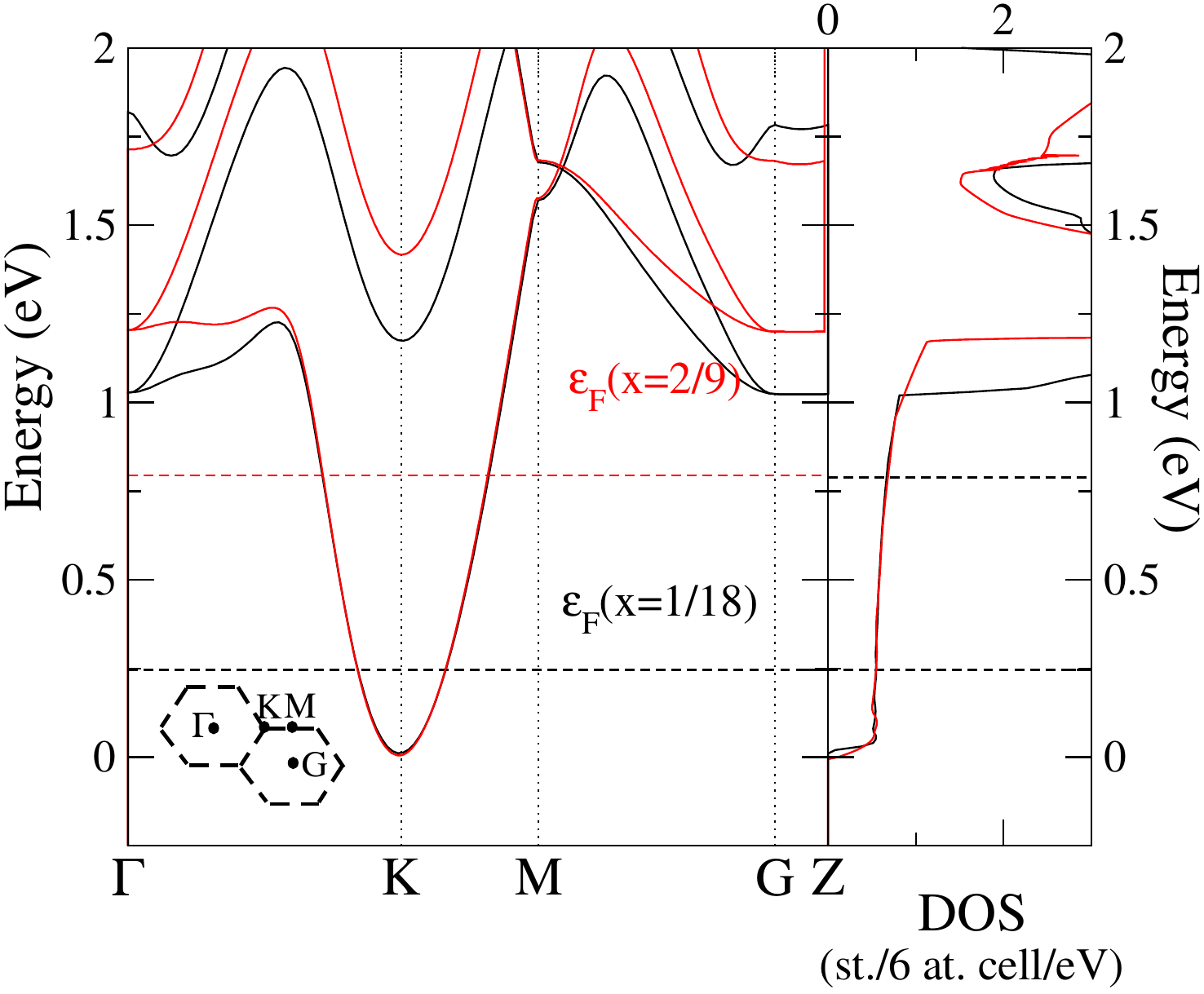}
\caption{Electronic structure and density of states of Li$_x$ZrNCl
for $x=1/18$ (black) and  $x=2/9$, (red).
}
\label{fig1}
\end{figure}

\begin{figure}
\includegraphics[width=0.46\textwidth]{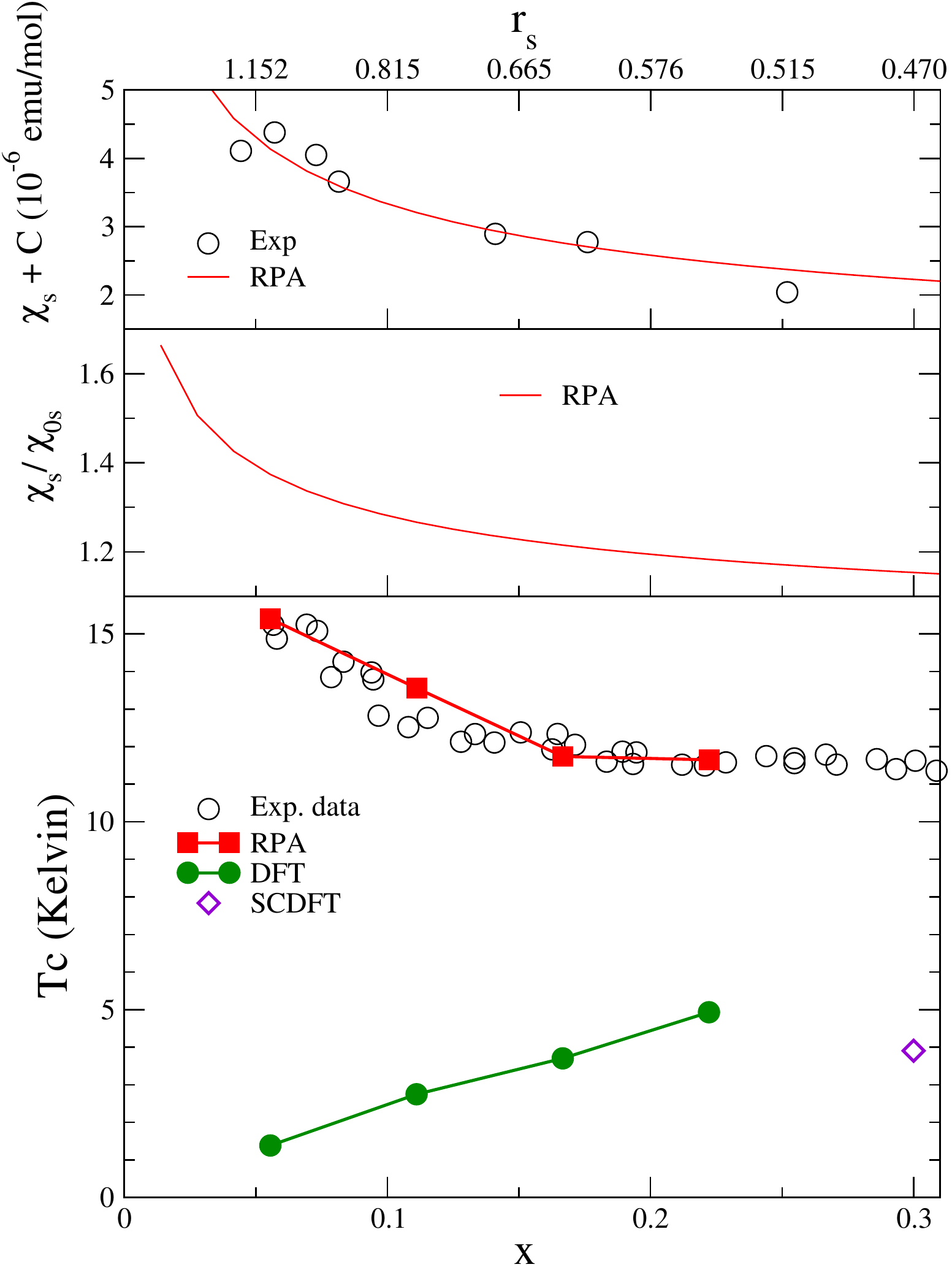}
\caption{Top: Spin susceptibility  ($\chi_{s}$) calculated in
  the random phase approximation (RPA) and experimental
  susceptibility. 
The experimental data 
from Ref. ~\cite{Kasahara} have been corrected
for an erroneous estimate of the Landau diamagnetic
susceptibility, $\chi_L$, (see
sec. A in \cite{SI} and supplementary materials in
Ref. ~\cite{Kasahara}). 
Center: RPA
enhancement factor ($\chi_s/\chi_{0s}$). Bottom:
Experimental~\cite{Kasahara}
and calculated $T_c$ using different
approximations. The
Superconducting Density Functional Theory calculation
is from Ref. \cite{Arita}  }
\label{fig3}
\end{figure}

ZrNCl is a layered large gap semiconductor, with an extremely weak
interlayer coupling ($t_{\perp}< 1.5\, {\rm meV}$) and 2 equivalent
valleys
with isotropic mass tensors in the conduction
band (see Fig. \ref{fig1} (a)) at the special points ${\bf K}$ and ${\bf K'}=2{\bf K}$.
The Li intercalation acts as
a rigid filling of the conduction band with $x$ electrons~\cite{Heid}.
The bands are almost parabolic with $m^{*}=0.57$ (in units of e mass) for doping $x\le 2/9$ (see Fig. \ref{fig1} (a)).
Li$_x$ZrNCl is thus a realisation of a 2D 2-valley
electron-gas.
The system remains insulating due to an
Anderson localization for $x \le 0.05$ and then
becomes superconducting at larger doping~\cite{Yamanaka}.
The spin-susceptibility $\chi_s$ increases as doping is reduced, as
shown in Fig.  \ref{fig3} (top-panel). As the non-interacting
$\chi_{0s}$ is doping independent in 2D, this increase can only
be due to exchange-correlation effects.
The superconducting $T_c$
behaves similarly to $\chi_s$ as it
{\it increases} from $11.5$ K to $15$ K
for $x$ {\it decreasing} from $0.3$ to $0.05$
(see Fig. \ref{fig3} bottom panel and
Ref. \cite{Taguchi}), suggesting that the two effects are related.

In order to evaluate the interacting $\chi_s$, we consider a 2D 2
valley electron-gas with a finite thickness equal
to that of the ZrN layer and
environmental dielectric constant
$\epsilon_M=5.59$,
as calculated density functional theory (DFT) for the
insulating compound ZrNCl~\cite{Galli}.
We obtain the $\chi_s/\chi_{0s}$
enhancement in the random-phase approximation (RPA)~\cite{SI}.
The RPA closely reproduces the quantum Monte-Carlo results~\cite{Zhang1, Zhang2, Marchi},
for $r_s$ values relevant for Li$_x$ZrNCl ($r_s < 1.5$).
As shown in Fig. \ref{fig3} central panel, the
enhancement is already large at these intermediate values of $r_s$.
To compare with measurements, we add to our $\chi_s$ a
constant $C$ that
takes into account the doping-independent 
Landau diamagnetic terms
(see Eq. 4 in \cite{SI}) present in the experimental data. 
Our $\chi_s$ closely reproduces the dependence on doping
measured in experiments~\cite{Kasahara,SI}.

In a Fermi liquid approach,
the electron-phonon coupling parameter
 of a mode $\nu$ at a phonon-momentum ${\bf q}$ is given by:
\begin{eqnarray}
\widetilde{\lambda}_{{\bf q}\nu} &=& \frac{2}{\omega_{{\bf q}\nu}^2N(0) N_{k}} \sum_{{\bf k}}
|\widetilde{d}_{{\bf k},{\bf k+q}}^{\nu}|^2 \delta(\epsilon_{{\bf k}})
\delta(\epsilon_{{\bf k+q}}) \label{eq:lambda}
\end{eqnarray}
where the tilde indicates that the quantities are fully screened by
all kind of exchange-correlation interaction (charge, spin and valley).
The quasiparticle energies
are $\epsilon_{{\bf k}} $ and
$\widetilde{d}_{{\bf k},{\bf k+q}}^{\nu}= \langle {\bf k}|\delta \widetilde{V}/\delta
u_{{\bf q}\nu} |{\bf k+q} \rangle$, with
$\widetilde{V}$ being the screened \cite{CalandraWannier} single-particle potential that includes, at the
mean-field level, the e-e interaction~\cite{footnote}.
Moreover $u_{{\bf q}\nu}$
and $\omega_{{\bf q}\nu}$ are the phonon displacement
and frequency.
In GGA or LDA functionals the exchange-correlation energy
depends explicitly on charge densities and spin polarization,
but not on valley polarizations. As a consequence
the SU$(4)$ spin and valley symmetry of the manybody Hamiltonian
is broken. Thus the matrix
elements in Eq. \ref{eq:lambda} do not
include any enhancement due to intervalley exchange-correlation~\cite{SI}.
They are then
undressed (bare) with respect to intervalley exchange-correlation
interaction. We label them as $d_{{\bf k},{\bf k+q}}^{\nu}$ and
${\lambda}_{{\bf q}\nu} $, without the tilde.
In the Hartree-Fock (HF) approximation, the
matrix elements include an intervalley exchange-correlation
enhancement, that is, however, severely overestimated with respect to
Quantum Monte Carlo or RPA results~\cite{Marchi}. For this reason, hybrid
functional calculations~\cite{Yin} lead to matrix elements
$\widetilde{d}_{{\bf k},{\bf k+q}}^{\nu}$ that crucially depends on
the amount
of HF exchange included.

\begin{figure}
\includegraphics[width=0.46\textwidth]{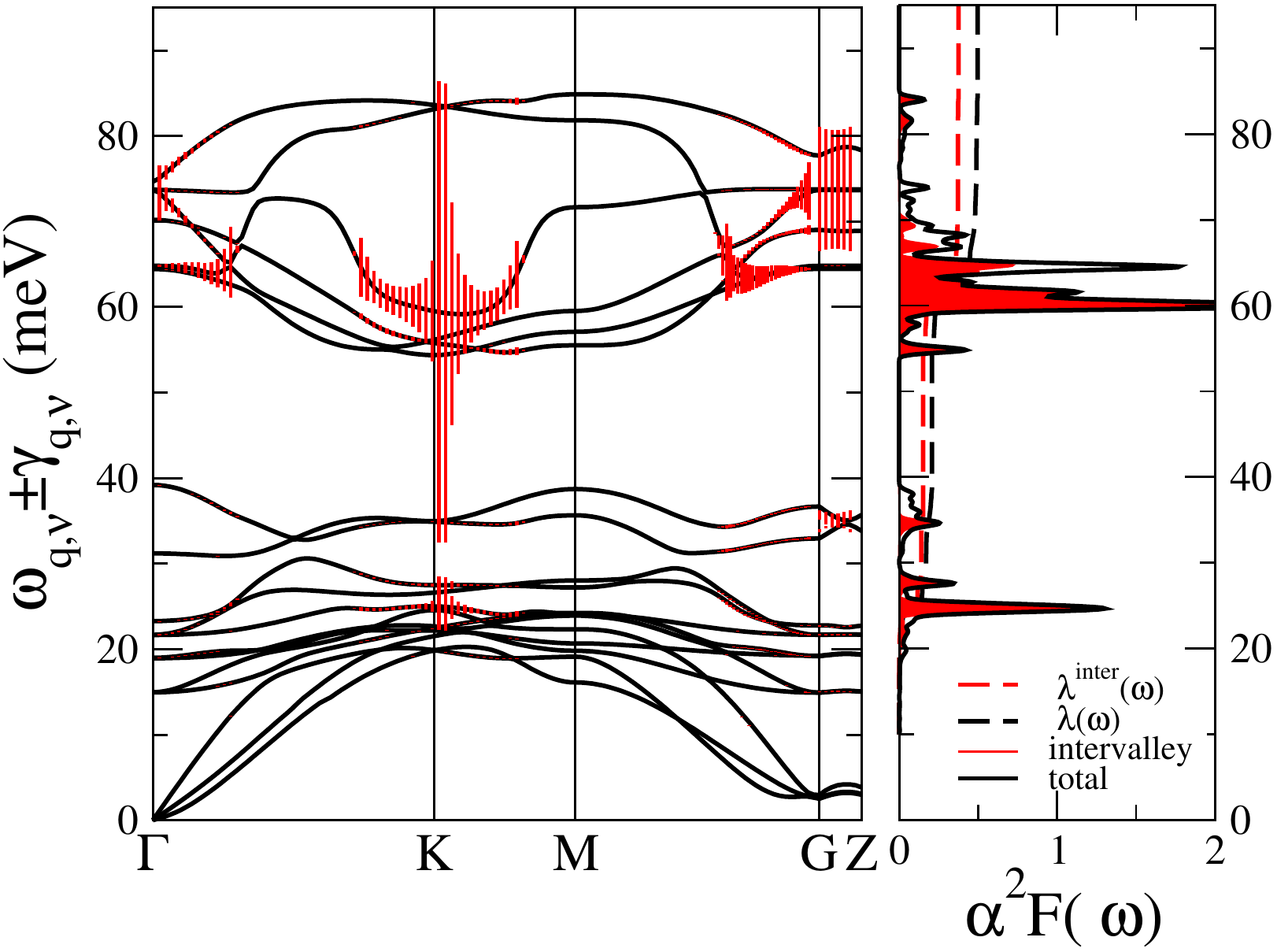}
\caption{Phonon dispersion, phonon linewidth (red bars) and Eliashberg
function of Li$_{1/18}$ZrNCl. The Eliashberg function due to
intervalley coupling only is shown as the filled region in the right
panel (see~\cite{SI} for other doping).}
\label{Fig:branchie}
\end{figure}

To evaluate the bare quanty
$\lambda_{{\bf q}\nu}$ as a function of doping, we use DFT~\cite{TecDet} and Wannier interpolation
 \cite{CalandraWannier}
(see~\cite{SI} for other doping).
We find a marked softening of
an intervalley phonon having $\approx 59$ meV phonon-energy
at $x=1/18$ in a
region of radius $2\kappa_F$ around ${\bf K}$ (Fig. \ref{Fig:branchie}).
As the softening $\Delta \omega_{{\bf
    q}\nu}$ is	essentially constant
in this region, we conclude that
$|d_{{\bf k},{\bf k+K}}^{\nu}|\approx |d_{{\bf K},{\bf 2K}}^{\nu}|$.
Indeed, under this assumption the phonon softening at ${\bf q}$ close
to ${\bf K}$ is
$\Delta \omega_{{\bf q}\nu}\approx -\chi_{0}({\bf q}) |d_{{\bf K},{\bf
    2K}}^{\nu}|^2/(2\omega_{{\bf q}\nu})$~\cite{SI}.
Here $\chi_{0}({\bf q})$ is
the bare finite-momentum response-function, which is constant and doping independent in 2D for
$|{\bf q}-{\bf K}|< 2\kappa_F$~\cite{GiulianiVignale,SI}.

The average electron-phonon coupling $\lambda$ as a function of doping
is shown in Tab. \ref{table1}.
We further decompose $\lambda$ in inter- and intra-valley
contributions finding that at low doping (i) the intravalley
contribution is suppressed for $x$ going to zero and
(ii) the intervalley contribution is almost
doping independent and
dominant, as shown in Fig. \ref{Fig:branchie} and in~\cite{SI}.
In the Eliashberg function
at $x=1/18$ most of the coupling arises
from intervalley phonons at $\approx 59$ and $24.5$ meV,
 (Fig. \ref{Fig:branchie}). These
phonons have large phonon linewidths
$\gamma_{{\bf q}\nu}=\pi N(0) \omega_{{\bf q}\nu}^2\lambda_{{\bf
    q}\nu}$,
as shown by the red bars in Fig. \ref{Fig:branchie}.
Finally, both the average electron-phonon coupling and the logarithmic average of the
phonon frequencies are roughly constant  (see Tab. \ref{table1} ), so that
$T_c$ as obtained from
McMillan equation \cite{mustar} slightly increases with doping, in agreement with
previous calculations at higher doping~\cite{Heid,Arita}, but in  qualitative
disagreement with experimental data (see Fig. \ref{fig3} bottom
panel).

Assuming a constant intravalley electron-phonon matrix element ($|d_{{\bf k},{\bf k+K}}^{\nu}|\approx |d_{{\bf K},{\bf 2K}}^{\nu}|$),  we can derive an
effective Hamiltonian where the presence of a small phonon
displacement $u_{{\bf K}\nu}$ is described as an external pseudo magnetic field
$B_{\rm ext}^{\nu} = |d_{{\bf K},{\bf 2K}}^{\nu}|u_{{\bf K}\nu}/\mu_S$.
Indeed, if we define a 2-component spinor using as basis the states
$|{\bf K}+\bm{\kappa}\rangle$ and $|2{\bf K}+\bm{\kappa}\rangle$,
where $\bm{\kappa}={\bf k}-{\bf K}$,
we obtain the following form of the one-body part of the Hamiltonian:
\begin{equation}
H^{\nu}_{\bm{\kappa}}=\frac{\hbar^2 \kappa^2}{2 m^*} \hat{I} + B_{\rm ext}^{\nu} \,\mu_S \hat{\sigma_x},
\end{equation}
where  $\hat{I}$ and $\hat{\sigma_x}$ are the $2\times2$ identity and the Pauli
 matrix along the x-direction, respectively.
Here,
without loss of generality, $B_{ext}^{\nu}$ has be chosen to be real by 
fixing appropriately the relative phase between the $|{\bf
  K}+\bm{\kappa}\rangle$ 
and $|{\bf 2K}+\bm{\kappa}\rangle$ states (see sec. G  in \cite{SI}).

\begin{figure}[t]
\begin{minipage}[l]{0.3\linewidth}
\hspace{-1.0cm}\includegraphics[width=3.5cm]{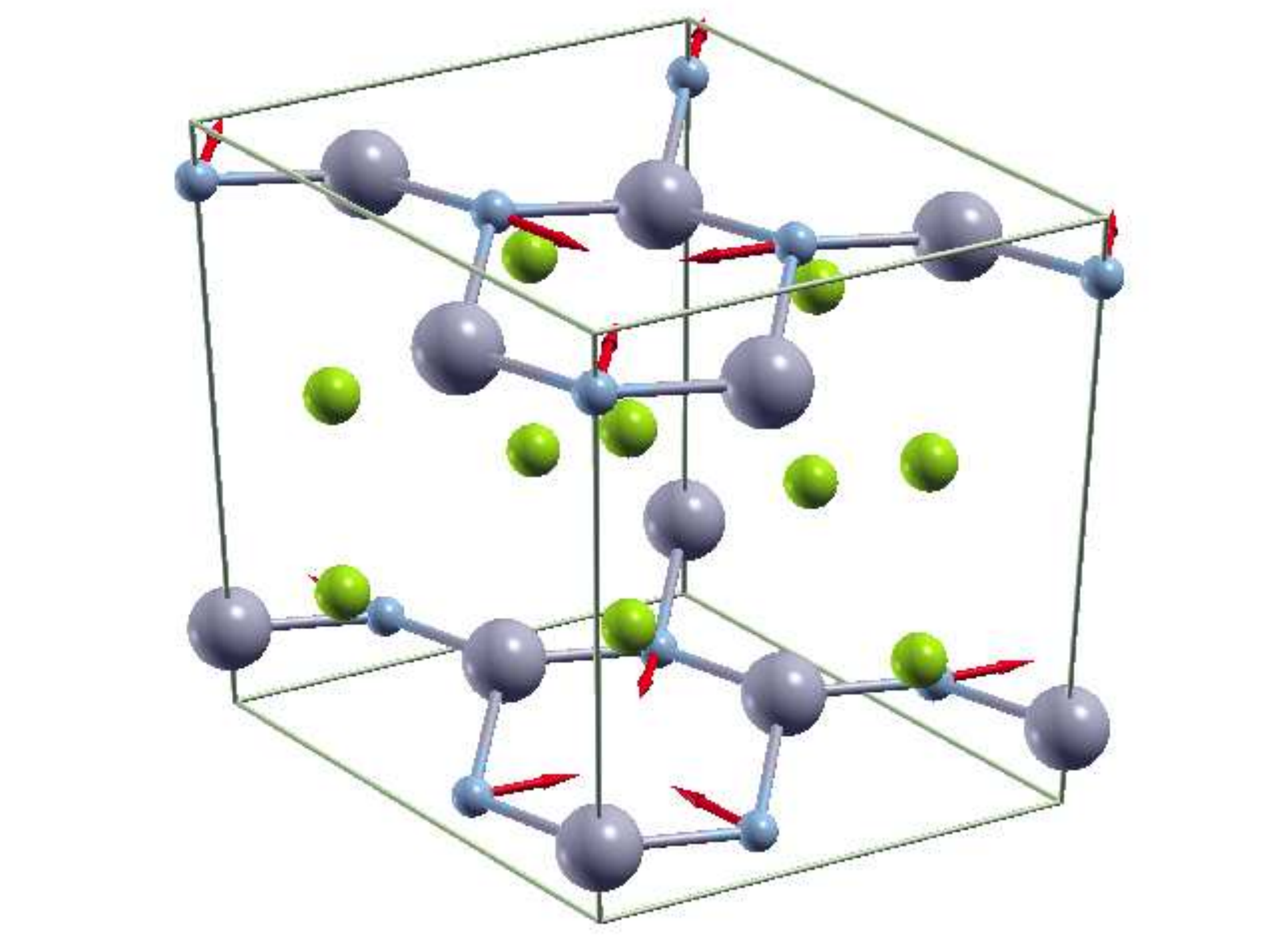}
\end{minipage}
\begin{minipage}[l]{0.5\linewidth}
\includegraphics[width=4.0cm]{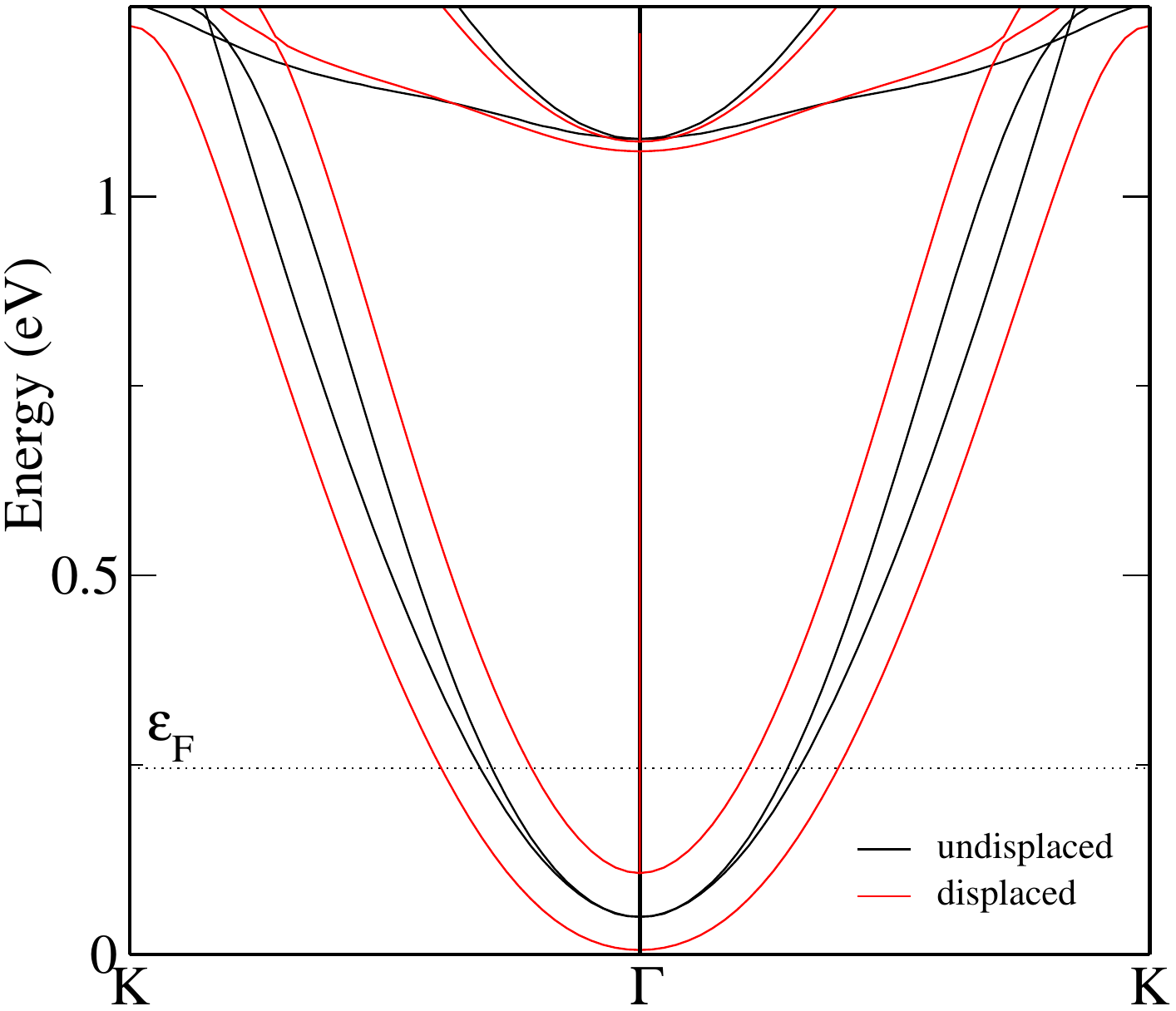}
\end{minipage}
\caption{(a) Phonon displacements of the $59$ meV modes at ${\bf K}$
in a $\sqrt{3}\times\sqrt{3}$ supercell with AA stacking. (b) Effect of
the phonon displacements in (a) on the electronic structure of the
$\sqrt{3}\times\sqrt{3}$ supercell with AA stacking. The dotted line
labels $\epsilon_F$. The point ${\bf K}$ is folded at ${\bf \Gamma}$
in the supercell. The displacements of the Zr and N atoms are $4\times
10^{-3}$\AA ~and $59\times 10^{-3}$\AA .}
\label{figsplit}
\end{figure}

We explicitly verify the accuracy of such Hamiltonian by performing a DFT electronic structure
calculations on a $\sqrt{3}\times\sqrt{3}$ supercell
with AA stacking. In this supercell, the 2 valleys at ${\bf K}$ and $2{\bf K}$ in the unit cell,
 are folded at the zone center. As shown in Fig. \ref{figsplit},
by displacing the
 atoms from the equilibrium, the intervalley
 phonon splits the 2 valleys by a constant amount equal to
 $2B_{\rm ext}^{\nu}\mu_S$, as predicted by the model Hamiltonian.
The intervalley phonons induce a valley polarization and act as a pseudo magnetic field.
As it happens in the magnetic case, the
response to the pseudo magnetic field is enhanced by the
intervalley exchange-correlation (which is however absent in our DFT
calculation, as shown in~\cite{SI}).
As the total magnetization due to the pseudo magnetic
field $B_{\rm ext}^{\nu}$ is written either as $M=\chi_s B_{\rm
  ext}^{\nu}$ or as $ M=\chi_{0s} \widetilde{B}^{\nu}$, where now $\widetilde{B}^{\nu}$
is the total magnetic field,
sum of the external plus the exchange-correlation field, we have~\cite{SI},
\begin{eqnarray}
\frac{\widetilde{B}^{\nu}}{B_{\rm ext}^{\nu}}=
\frac{|\widetilde{d}_{{\bf K},{\bf 2K}}^{\nu}|}{|d_{{\bf K},{\bf
 2K}}^{\nu}|}=\frac{\chi_s}{\chi_{0s}}
\label{eq:drenK}
\end{eqnarray}
namely the electron-phonon coupling at ${\bf q}={\bf K}$ is renormalized by intervalley correlation
effects exactly in the same way as the spin susceptibility with an enhancement that is independent from the phonon index $\nu$.
Assuming again a constant intervalley matrix element we have that:
\begin{eqnarray}
{\tilde \lambda}^{\rm inter}= \left(\frac{\chi_s}{\chi_{0s}}\right)^2
\lambda^{\rm inter}
\label{eq:lamtinter}
\end{eqnarray}
so that ${\tilde \lambda}=\lambda^{\rm intra}+{\tilde \lambda}^{\rm inter}$.
We use the $\chi_s/\chi_{0s}$ of Fig. \ref{fig3} (central
panel)
to evaluate ${\tilde  \lambda}$,
the renormalized Eliashberg function and T$_c$ using
McMillan equation \cite{mustar}. The results are shown in Fig.  \ref{fig3} (bottom
panel).
We now find that the doping dependence of 
T$_c$ is in excellent agreement with
experimental data. In addition, by using 
a reasonable value of $\mu$ we also obtain T$_c$ in agreement with
experiments.
Intervalley
exchange-correlation is the mechanism responsible for the enhancement
of T$_c$ at low doping in Li$_x$ZrNCl.

Here we have shown that the
e-e interaction enhances T$_c$ at low doping in Li$_x$ZrNCl.
Such finding is universal and provides a
general guideline to realize a superconducting state in a doped semiconductor.
First of all the system should be strongly 2D . In this
case the density of states is doping independent and constant down
to very low doping, where Anderson localization occurs. In
three dimensional multivalley doped semiconductors, the enhancement
of the valley susceptibility due to manybody effects still occurs, 
but the density of states tends to zero at low doping and T$_c$ is
suppressed \cite{Ekimov,Bustarret,Ren,Kriener}.
Furthermore, in order for the enhancement to occur, a multivalley electronic
structure is needed but 2 is the optimal number of
valleys. Indeed, the enhancement is smaller as the number
of valleys increases and ultimately tends to one in the limit of
infinite number of valleys. 
Finally, the T$_c$ enhancement is larger, the larger the
$r_s$ parameter. A larger $r_s$ parameter can be obtained reducing
the doping, reducing the dielectric constant of the
spacers ($\epsilon_M$) or increasing $m^*$.
At fixed $\epsilon_M$ and $m^*$,
the largest enhancement should be found in the proximity of the
band insulating or semiconducting state. In the very low doping
limit, in the absence of disorder, the enhancement of T$_c$ can be so
large to induce high T$_c$ superconductivity. However,
in this limit, disorder and the resulting Anderson localization tend
to suppress superconductivity. Thus, high T$_c$ superconductivity
will only be seen in very clean samples.
Finally, it is worth to recall than the pairing mechanism does not need
to be necessary the electron-phonon interaction. Indeed, any mechanism
(e.g. spin-fluctuations) inducing a valley polarization will
experience an enhancement of T$_c$ due to intervalley
exchange-correlation.

\begin{table}[h]
\caption{$\epsilon_F$, $\omega_{\log} $, $\lambda$,
$\omega_{\log}^{\rm intra}$, $\lambda^{\rm intra}$ ,
$\omega_{\log}^{\rm inter}$ ,$\lambda^{\rm  inter}$  from 
DFT. Energies in meV.
The enhancement $\chi_v/\chi_{0v}$ is calculated in
the RPA approximation. The couplings
$\widetilde{\lambda}^{\rm inter}$ and $\widetilde{\lambda}$ are
obtained via Eq. \ref{eq:lamtinter}.}
\begin{center}
\begin{tabular}{l cc c c c c c c c c c} 
\hline
\hline
$x $ & $\epsilon_F $ & $\omega_{\log} $ & $\lambda$ & $\omega_{\log}^{\rm intra}$ &
$\lambda^{\rm intra}$ & $\omega_{\log}^{\rm inter}$ & $\lambda^{\rm
  inter}$ &
$\chi_v\over\chi_{0v}$ & $\mu^*$ & ${\tilde
  \lambda}$ \\
\hline
1/18   & 246 &  43.9  &  0.48    &  48.0 &  0.10 &  42.8 &
0.38  & 1.37 & 0.165 &  0.82  \\ 
1/9    & 448 &   44.8 & 0.51     &  47.6 &  0.11 &  43.9 &
0.39  & 1.27 & 0.150 & 0.74 \\ 
1/6    & 622 &   44.6  &  0.52    &  45.5 &  0.16 &  44.3
& 0.36 & 1.21 & 0.143 & 0.69   \\
2/9    & 790 &   43.3  &  0.55   &  42.9 &  0.20 &  43.4 &
0.35  & 1.18 & 0.138 & 0.69  \\
\hline
\hline
\end{tabular}
\end{center}
\label{table1}
\end{table}

%
%

We acknowledge discussions with M. L. Cohen ,S. de Palo, R.Heid, M. Johannes,
I. I. Mazin and S. Moroni, and 
support from the Graphene Flagship and
by ANR-11-BS04-0019 and ANR-13-IS10-0003-01.
Computer facilities were provided by CINES, CCRT and IDRIS.




\newpage
\large
\onecolumngrid
\subsection{\large Supplementary Materials of \\
{\it  Weakly-doped two-dimensional semiconductors: a new route towards
high T$_c$ superconductivity.  }}

\subsection{\large Spin susceptibility from experiments}

Experiments do not measure directly the
spin-susceptibility of conduction electrons, but the sum of orbital
and spin contributions of {\it all} the electrons present in the system.
The contributions of non-conducting electrons and the orbital
contribution of conducting electrons (Landau susceptibility, see
discussion below)  are doping independent. Thus, the raw experimental susceptibility data 
cannot be compared directly with the spin susceptibility of conduction electrons.

For this reason in Ref. \cite{Kasahara} (see, in particular, the discussion in the supplementary information of Ref. \cite{Kasahara}) 
all possible diamagnetic contributions where subtracted
from the
measured susceptibility $\chi$, namely:
\begin{equation}
\chi_s=\chi-\chi_{\rm core}^{\rm Li^{+}} -\chi_{\rm core}^{\rm ZrNCl}
-\chi_L -\chi_{orb}
\label{all:iwasa}
\end{equation}
where $\chi_{\rm core}^{\rm Li^{+}} $ and $\chi_{\rm core}^{\rm
  ZrNCl}$ are the core diamagnetic susceptibility from Li ion and
pristine $\beta-$ZrNCl, respectively. The quantities $\chi_L$ and
$\chi_{\rm orb}$ are the Landau diamagnetic and orbital
susceptibilities, respectively. 
The $\chi_{\rm core}^{\rm Li^{+}} $ and $\chi_{\rm core}^{\rm ZrNCl}$
susceptibilities are doping independent.
The orbital susceptibility $\chi_{\rm orb}$ was considered zero for a magnetic
field applied along the $c$ direction.
Finally, in Ref. \cite{Kasahara} the Landau susceptibility  $\chi_L$
was assumed to be
\begin{equation}
\chi_L=-\frac{1}{3 m^*}\chi_s 
\label{eq:chiLiwasa}
\end{equation}
where $m^*$ is the band
effective mass in units of the
electron mass ($m^*=0.66$ in \cite{Kasahara}). This last assumption is not correct. Indeed,
in a 2D electron gas with parabolic bands:
\begin{equation}
\chi_{0L}=-\frac{1}{3(m^*)^2 }\chi_{0s}
\end{equation}
where $\chi_{0L}$ is the non interacting Landau susceptibility.
Moreover, 
it has been shown that many-body effects
strongly renormalize the spin susceptibility, while the Landau
susceptibility is only weakly renormalized~\cite{Vignale, VignalePRL}.
For the doping regime considered here we can assume that $\chi_{0L}=\chi_{L}$.
Thus Eq.~\ref{eq:chiLiwasa} should be replaced by:
\begin{equation}
\chi_L=-\frac{1}{3(m^*)^2 }\chi_{0s}=-\frac{\mu_S^2}{3(m^*)^2}N(0).
\label{eq:iwasacorrect}
\end{equation}

Since we are only interested in the variation of the susceptibility with 
doping, and $\chi_{0s}$ is doping independent, 
we add back to the experimental data the negative quantity, erroneously removed with Eq.~\ref{eq:chiLiwasa}. 
This is done by multiplying the susceptibilities presented in Fig.~4 of Ref.~\cite{Kasahara} by a $[1-1/(3m^*)]=0.495$ factor.  The results are reported in Fig. 2 in our main paper as measured data.

\subsection{\large Model Hamiltonian and SU$(2g_v)$ spin-valley symmetry in the low density limit}

We consider an isolated band partially filled with electrons. 
Within this band, the electrons experience
a Coulomb repulsion
\begin{equation}
v(q)=\frac{2 \pi e^2}{\epsilon_M q}
\end{equation}
where ${\bf q}$ is the exchanged momentum between the two interacting electrons. 
The effect of the screening of other (empty) conduction and (filled)
valence bands is included via the environmental
dielectric constant $\epsilon_M$.
We can define two types of electron-electron scattering: i) the
intravalley scattering with $q\sim\kappa_F$ (
$\kappa_F$ being the Fermi momentum measured from the valley bottom), that does not change the  valley index of the electrons, ii) the 
intervalley scattering with $q\sim|{\bf K}-{\bf K}^{\prime}|=|{\bf K}|$ (${\bf
 K}$ and ${\bf K^{\prime}}=2{\bf K}$ being the positions of the valley
bottoms in the Brillouin zone), that changes the  valley index of the electrons.

In the low doping limit, namely for $\kappa_F\ll |{\bf K}-{\bf K}^{\prime}|$,
because of the divergence of the Coulomb repulsion for $q\to 0$, the
intravalley scattering is dominant and the intervalley scattering can be neglected.
Under this hypothesis, the valley index (as the spin index) is conserved by the Coulomb interaction, it can be treated as a pseudospin and
the manybody Hamiltonian has exact SU$(2 g_v)$
spin and valley symmetry, namely
\begin{equation}
H=\sum_{{\bm \kappa} v \sigma}\frac{\hbar^2 \kappa^2}{2 m^*}
c^{\dagger}_{{\bm \kappa} v\sigma} c_{{\bm \kappa} v\sigma} +
\sum_{{\bm \kappa} v\sigma}
\sum_{{\bm \kappa^{\prime}} v^{\prime} \sigma^{\prime}}
\sum_{\bm q} v(q) 
c^{\dagger}_{{\bm \kappa} v\sigma} 
c^{\dagger}_{{\bm \kappa^{\prime}} v^{\prime} \sigma^{\prime}}  
 c_{{\bm \kappa^{\prime}}-{\bm q} v^{\prime}\sigma^{\prime}} 
 c_{{\bm \kappa}+{\bm q} v\sigma} 
\label{Eq:H_ando}
\end{equation}
where $v,v^{\prime}=1,...,g_v$ are valley indexes and $\sigma,
\sigma^{\prime}=\pm$ are spin indexes and $c$,$c^{\dagger}$ are
creation and destruction operator (see e. g. Eq. 3.35 of
Ref. \cite{Ando}).
The Hamiltonian in Eq. \ref{Eq:H_ando} holds as long as
(i) the screening of the other bands can be included in the environmental
dielectric constant, (ii) intervalley scattering can be neglected. If
these two conditions are satisfied, then it holds regardless of the
number of valleys and of their position in the Brillouin zone.

As the Hamiltonian in Eq. \ref{Eq:H_ando} has exact SU$(2 g_v)$
spin and valley symmetry, it follows that:
\begin{equation}
\chi_v = \chi_s
\end{equation}

\subsection{\large Spin susceptibility in the random phase approximation}

We compute the interacting spin susceptibility of a multivalley 2D electron gas in the random phase approximation.
We integrate numerically the expression given by \cite{Zhang1}, which
(after correcting few typos, namely, in \cite{Zhang1}, the $1$ present on the r.h.s. of 
our equation below is missing 
and both the definitions of $A$ and $B$ are incorrect), reads:
\begin{eqnarray}
\frac{\chi_{0s}}{\chi_s} &=& 1-\frac{2\alpha r_s}{\pi}\int_{0}^{1} dx \frac{x
F(x)}{\sqrt{1-x^2}}
+\frac{\sqrt{2}\alpha r_s}{\pi}\int_{0}^{\infty} x^2 F(x) dx
\int_{0}^{\infty} du \left[\frac{1}{\epsilon(x,iu)} -1\right] \nonumber
\\
&\times& \left(A\sqrt{1+A/R} - B\sqrt{1-A/R}\right) R^{-5/2}
\label{eq:chisuchi0}
\end{eqnarray}
where $\alpha=\sqrt{g_v g_s /4}$ with $g_v$ ($g_s$) being the valley
(spin) degeneracy, $x=q/2k_F$, $A=x^4 -x^2 -u^2$, $B=2 x^2 u$, 
$R=\sqrt{A^2+B^2}$ and $u=\omega/(4 \epsilon_F)$.
The imaginary frequency dielectric function $\epsilon(x,iu)$ is defined as:
\begin{eqnarray}
\epsilon(x,iu)=1+\alpha r_s g_v g_s F(x) \left[\frac{1}{2x} -
\sqrt{A+R/(2^{3/2} x^3)}\right]
\end{eqnarray}
The wavefunctions of conduction bands are localized on ZrN bilayers.
The thickness of each bilayer (distance between Zr and N along the
z-axis) is $\approx 2.15 \AA$. Considering the extension of the DFT
charge density we set  $a=2.5$~$\rm \AA $. 

In addition we also consider the perfect (long-range) metallic screening of the adjacent ZrN bilayers, located at a distance $d=9.306$~$\rm \AA$.
We encode this information in the form factor:
\begin{eqnarray}
F(x)=
\frac{2}{q a^*}(1+\frac{e^{-q a^*}-1}{q a^*})
+\frac{1-e^{-4 q d^*}}{1+e^{-4 q d^*}+2 e^{-2 q d^*}} - 1
\label{eq:form_factor}
\end{eqnarray}
where  $q=2xk_F$, $k_F=1/(r_s \alpha)$,
$a^*=a/a_B$, $d^*=d/a_B$, $a_B=(\epsilon_M /m^*) 0.529177$~${\rm \AA}$, and 
we suppose that $d \gg a$. 

\begin{figure}[h]
\includegraphics[width=0.8\textwidth]{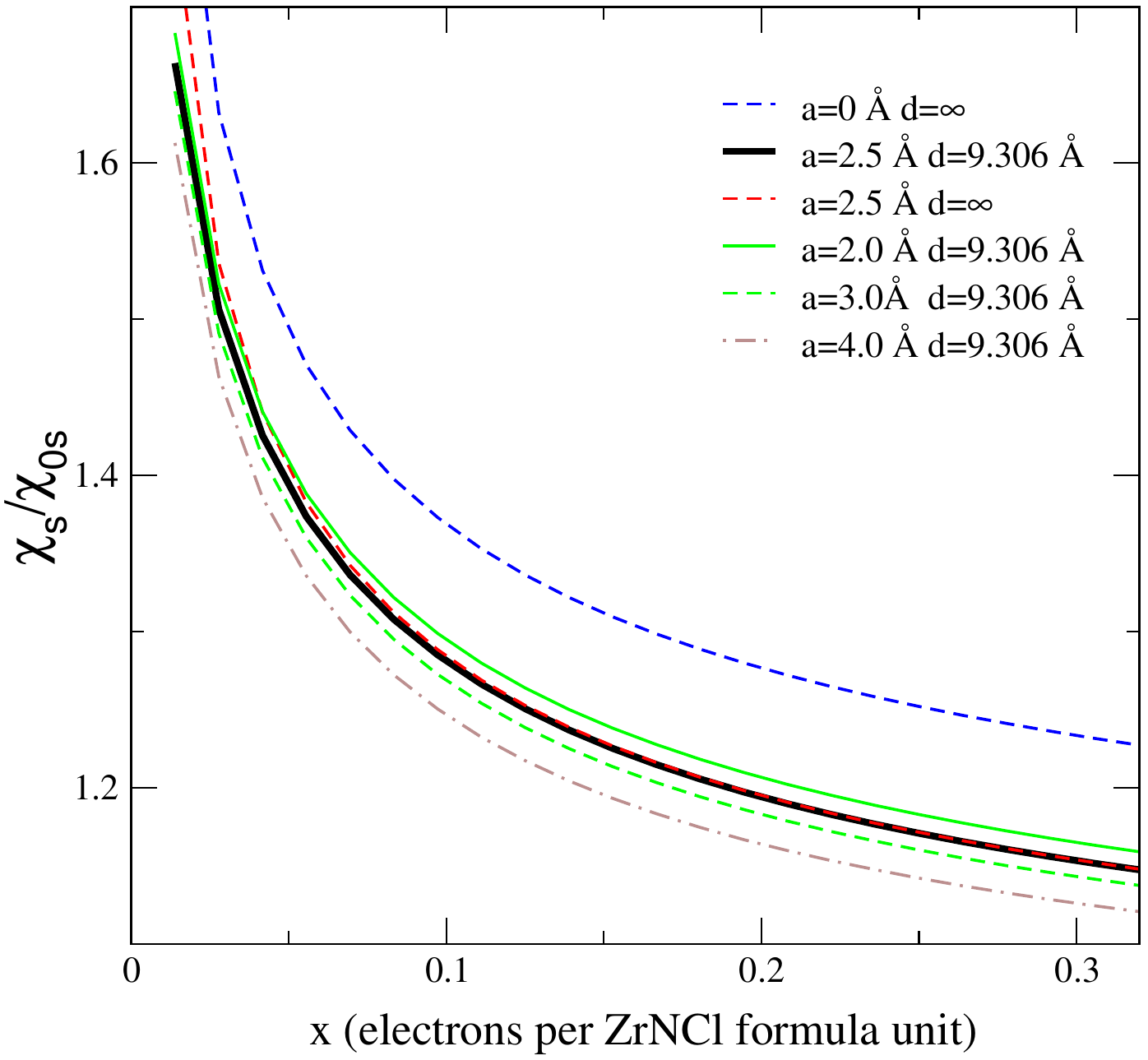}
\caption{
Susceptibility enhancement as a function finite thickness $a$ of the
2D electron-gas and of the interlayer distance $d$. In the main paper we report the results obtained with  $a$=2.5 \AA~and $d$=9.306 \AA. 
}
\label{figs_chi_enhancement}
\end{figure}
The result of Eq. \ref{eq:chisuchi0} are shown in
Fig. \ref{figs_chi_enhancement}. In the same picture we also compare
the effect of the parameter  $a$ 
and of the metallic screening of the adjacent ZrN bilayers. The
presence ($d=9.306\AA$) or absence ($d=\infty$) of metallic screening has no influence on the
ratio $\chi_s/\chi_{0s}$. Furthermore, 
the dependence on thickness is extremely weak for realistic values of
$a$, namely $2 < a < 3$. The choice of this parameter is thus not critical.

In the top panel of Fig.~2 of our main paper, we add a doping independent constant C to the RPA result for $\chi_s$
to account for the Landau diamagnetic term, Eq.~\ref{eq:iwasacorrect}, 
and for the uncertainties on the estimations of the others diamagnetic (doping independent) terms of the left hand side of Eq~\ref{all:iwasa}. 
The best agreement with experiment is obtained for C$=-7.77 \times 10^{-6}$~emu/mol. Note that, using a value of $m^*=0.57$ as in the rest of the paper, Eq.~\ref{eq:iwasacorrect} gives $\chi_L=-8.89\times 10^{-6}$~emu/mol, in close agreement with the value obtained for the constant C.

\subsection{\large Spin susceptibility in local spin density functional theory}

The total energy in local spin density functional theory in the
presence of an external (bare) magnetic field $B_{\rm ext}$ is written as:
\begin{equation}
E_{LSD}=T+\int d{\bf r}\, \epsilon_{h,xc}(n,m) - \mu_s \int d{\bf r}\,
m({\bf r}) B_{\rm ext}
\end{equation}
where $T$ is the kinetic energy functional, 
$\epsilon_{h,xc}(n,m)=\epsilon_{h}(n)+\epsilon_{xc}(n,m)$ is the
Hartree and exchange and correlation energy per particle, 
$n({\bf r})$ is the electron density and 
$m({\bf r})=n_{+}({\bf r})-n_{-}({\bf r})$.
The Kohn-Sham potential for each spin channel ($\pm$) is written as
\begin{equation}
V_{KS}^{\pm}=\frac{\delta \epsilon_{h,xc}}{\delta n({\bf r})} \pm
\frac{\delta \epsilon_{h,xc}}{\delta m({\bf r})}\mp \mu_s B_{\rm ext} 
\end{equation}
In a paramagnetic system\cite{Gunnarsson}, we have that
\begin{equation}
\left.\frac{\delta \epsilon_{xc}(n,m)}{\delta m}\right|_{m=0}=0
\end{equation}
We can then expand at second order $\epsilon_{xc}(n,m)$ in $m$ and
obtain
\begin{equation}
V_{KS}^{\pm}=\frac{\delta \epsilon_{h,xc}}{\delta n({\bf r})} \pm
\frac{\delta^2 \epsilon_{xc}}{\delta m({\bf r})^2}\, m\mp \mu_s B_{\rm ext} .
\end{equation}
We call 
\begin{equation}
\widetilde{B}=B_{\rm ext}+B_{xc} \label{def1}
\end{equation}
the total (screened by the exchange-correlation) field, where
\begin{equation}
\mu_s B_{xc}=-\frac{\delta^2 \epsilon_{xc}}{\delta m({\bf r})^2}\, m.\label{def2}
\end{equation}
The 
fields are related to the
magnetization by the following two relations:
\begin{eqnarray}
\mu_s m &=& \chi_{0s} \widetilde{B} \label{eq:chi0B},\\
\mu_s m &=& \chi_{s} B_{\rm ext} \label{eq:chiB}.
\end{eqnarray}  
where $ \chi_{0s}$ and $\chi_{s}$ are the bare and interacting
susceptibilities, respectively.
Combining Eq.s~\ref{def1}, \ref{def2}, and \ref{eq:chiB}, we obtain:
\begin{equation}
\widetilde{B}=(1-f \chi_s) B_{\rm ext}
\label{eq:BvsBext}
\end{equation}
where $f=\frac{\delta^2 \epsilon_{xc}}{\delta m({\bf
    r})^2}\frac{1}{\mu_s^2} < 0$. From Eqs. \ref{eq:chi0B} and
\ref{eq:chiB} we have 
\begin{equation}
\frac{\widetilde{B}}{B_{\rm ext}} = \frac{\chi_s}{\chi_{0s} }
\label{eq:BoverB}
\end{equation}
meaning that the total magnetic field in the sample is renormalized by
exchange-correlation effects exactly as the manybody enhancement of
the spin susceptibility.

Replacing Eq. \ref{eq:BoverB} in Eq. \ref{eq:BvsBext}, we obtain:
\begin{equation}
\frac{\chi_s}{\chi_{0s}}=\frac{1}{1+f\chi_{0s}} \label{eq:BBext}
\end{equation}
The relation between the total (screened) magnetic field and the
external (bare) one is determined by the function $f$ that is related
to the second derivative of the exchange correlation potential with
respect to the spin magnetization. 
If the exchange correlation functional has
no dependence on $m$, then $f=0$ and there is no susceptibility enhancement.

\subsection{\large Absence of valley susceptibility enhancement in local spin density}
In Li$_x$ZrNCl 
a valley polarization
can be induced by the atomic displacements of an intervalley phonon, 
as shown in Fig. 4 in the main paper.
In this case the deformation potential acts as an external (bare) 
$B_{\rm ext}^{\nu} = |d_{{\bf K},{\bf 2K}}^{\nu}|u_{{\bf K}\nu}/\mu_S$
pseudo magnetic field.

In an exact many-body treatment, in the low doping limit, the $SU(4)$ spin and valley
symmetry is preserved. This is not
necessary the case if approximated local exchange and correlation functionals are used. 
If a 4-component local exchange correlation kernel is
adopted in the calculation, namely
\begin{equation}
\epsilon_{xc}=\epsilon_{xc} (n, m, m_v)
\end{equation}
where $m_v({\bf r})=n_{v=1}({\bf r})-n_{v=2}({\bf r})$ is the valley
magnetization and $\epsilon_{xc} (n, m, m_v)=\epsilon_{xc} (n, m_v,
m)$, then the $SU(4)$ symmetry can be preserved.
The  valley exchange-correlation enhancement is written as
\begin{equation}
\frac{\chi_s}{\chi_{0s}}=\frac{1}{1+f_v\chi_{0s}}
\end{equation}
with $f_v= \frac{\delta^2 \epsilon_{xc}}{\delta m_v({\bf
    r})^2}\frac{1}{\mu_s^2}<0 $.
Thus if the exchange and correlation energy per particle depends
explicitly on $m_v$, there is a valley exchange-correlation enhancement different
from $1$.

In standard local LDA/GGA functionals, the $SU(4)$ spin valley 
symmetry is broken and
the exchange and correlation energy per particle is assumed to be
\begin{equation}
\epsilon_{xc}=\epsilon_{xc} (n, m)
\end{equation}
independent of $m_v$. In this case, $f_v=0$ and the valley exchange-correlation
enhancement is exactly one. Thus the valley susceptibility is bare in
this case.

Finally, it is important to remark that standard LDA/GGA
parametrizations of the electron gas are for the three dimensional
case. Thus the exchange-correlation enhancement of spin and valley
susceptibilities is taken into account incorrectly.

\subsection{\large Relation between phonon softening and bare susceptibility}

The softening at ${\bf q}$ in Fermi liquid
theory with an effective single particle potential
\cite{CalandraWannier} is written as:
\begin{equation}
\widetilde{\Delta \omega}_{{\bf q},\nu}=\frac{1}{N_k}\sum_k\frac{|\widetilde{d}_{{\bf k},{\bf
     k+q}}^{\nu}|^2}{\omega_{{\bf q},\nu}} \frac{f_{{\bf k}+{\bf
     q}} - f_{{\bf k}}}{\epsilon_{{\bf k+q}}-\epsilon_{{\bf k}}}\label{eq:softening}\\
\end{equation}
the tilde means screened with respect to intervalley exchange correlation.
Assuming a constant intervalley matrix element ($|d_{{\bf k},{\bf
    k+K}}^{\nu}|\approx |d_{{\bf K},{\bf 2K}}^{\nu}|$),
we have for the phonon softening at phonon momentum ${\bf q}$ close to
${\bf K}$:
\begin{equation}
\widetilde{\Delta \omega}_{{\bf q},\nu}=-
\frac{|\widetilde{d}_{{\bf K},{\bf 2K}}^{\nu}|^2}{2\omega_{{\bf q},\nu}}\chi_{0}({\bf q}),\label{softconstant}
\end{equation}
where the bare finite-momentum response-function is
\begin{equation}
\chi_{0}({\bf q})=
-\frac{2}{N_k}\sum_k
     \frac{f_{{\bf k}+{\bf
     q}} - f_{{\bf k}}}{\epsilon_{{\bf k+q}}-\epsilon_{{\bf k}}}\label{eq:soften
ing}\\.
\end{equation}
For the parabolic 2-valley band-structure of Li$_{x}$ZrNCl ad ${\bf
  q}={\bf K}$ we have:
\begin{equation}
\widetilde{\Delta\omega_{{\bf K},\nu}}
 =
-\frac{|\widetilde{d}_{{\bf K},{\bf
   2K}}^{\nu}|^2}{2\omega_{{\bf K}\nu}\mu_S^2}\chi_{0s} 
\label{eq:gchibis},
\end{equation}
since $\chi_{0}({\bf K})=\lim_{{\bf q} \to {\bf K}} \chi_{0}({\bf q})
=N(0)=\chi_{0s}/\mu_S^2$.

\subsection{\large Electron-phonon interaction as a pseudo-magnetic field}

In the basis 
$\left\{|{\bf K}+{\bm \kappa}\rangle, |{\bf 2K}+\bm{\kappa}\rangle\right\}$,
the most general Hamiltonian that describes the coupling with an intervally phonon of momentum ${\bf K}$ and branch index $\nu$ is:
\begin{equation}
H_{\rm el-ph}=\left(\begin{array}{cc}
0 & d_{{\bf K}+{\bm\kappa},{\bf
    2K}+\bm{\kappa}}^{\nu}u_{{\bf K}\nu}\\
(d_{{\bf K}+\bm{\kappa},{\bf
    2K}+\bm{\kappa}}^{\nu})^* u_{{\bf K}\nu}& 0 \\
\end{array}
\right)
\end{equation}
where $d_{{\bf K}+\bm{\kappa},{\bf
    2K}+\bm{\kappa}}^{\nu}$ is the deformation potential and $u_{{\bf K}\nu}$ the amplitude of the phonon displacement.
By writing
\begin{equation}
d_{{\bf K}+\bm{\kappa},{\bf 2K}+\bm{\kappa}}^{\nu}
=|d_{{\bf K}+\bm{\kappa},{\bf 2K}+\bm{\kappa}}^{\nu}|e^{i\theta(\bm{\kappa})},
\end{equation}
we can obtain a real Hamiltonian by a redefinition of the basis,
$|{\bf 2K}+\bm{\kappa}\rangle \mapsto e^{-i\theta(\bm{\kappa})} |{\bf 2K}+\bm{\kappa}\rangle $. In this new basis:
\begin{equation}
H_{\rm el-ph}=\left(\begin{array}{cc}
0 & |d_{{\bf K}+\bm{\kappa},{\bf
    2K}+\bm{\kappa}}^{\nu}|u_{{\bf K}\nu}\\
|d_{{\bf K}+\bm{\kappa},{\bf
    2K}+\bm{\kappa}}^{\nu}| u_{{\bf K}\nu}& 0 \\
\end{array}
\right).
\end{equation}
Finally, since intervalley matrix element is
constant at small $\bm{\kappa}$, in the limit of small doping, we can ignore the $\bm{\kappa}$ dependence to obtain Eq. (3) of the main paper.

\subsection{\large Phonon dispersion in Li$_{x}$ZrNCl as a function of
  doping}

\begin{figure}[h]
\includegraphics[width=0.46\textwidth]{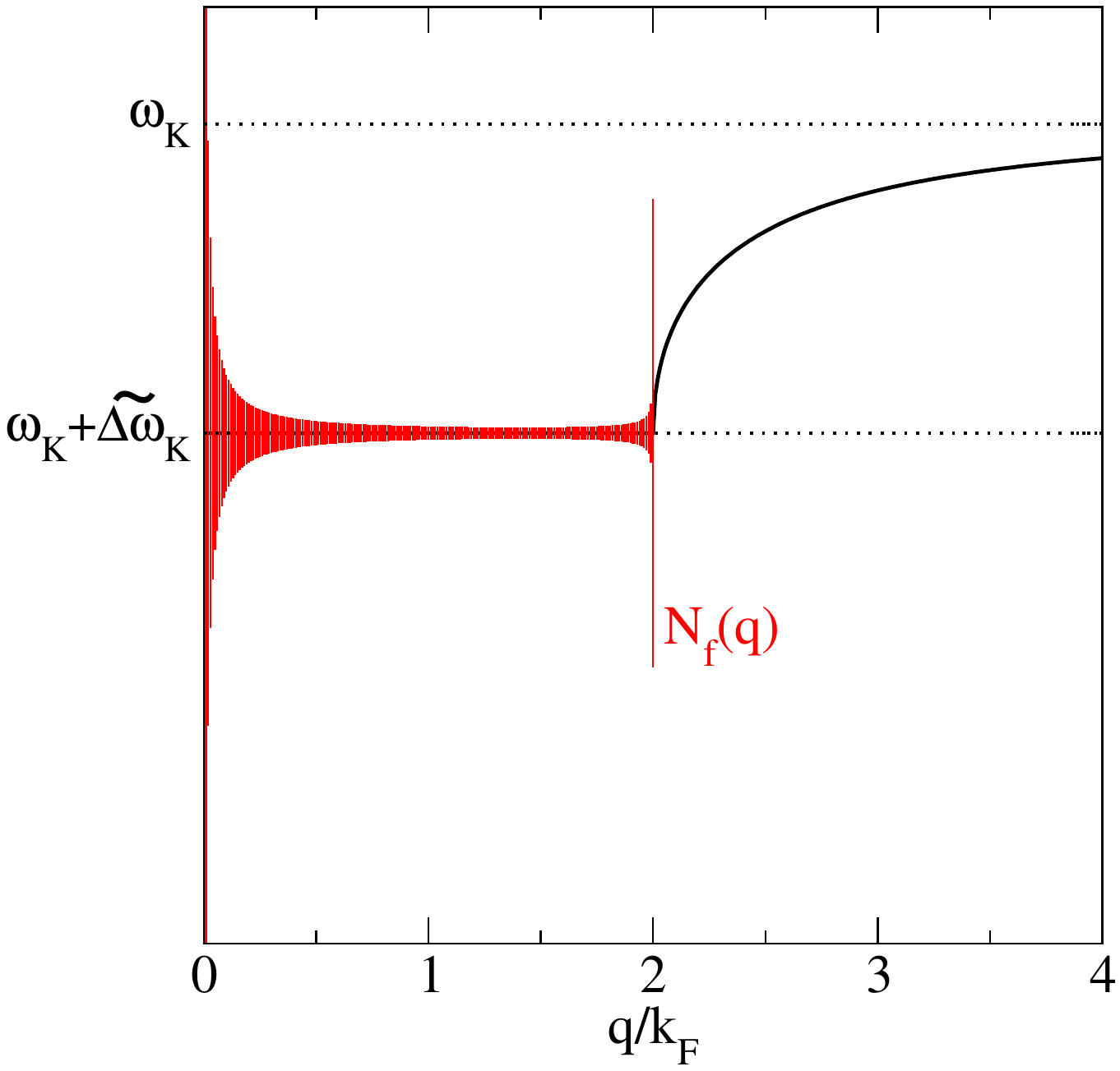}\hspace{0.5cm}
\includegraphics[width=0.47\textwidth]{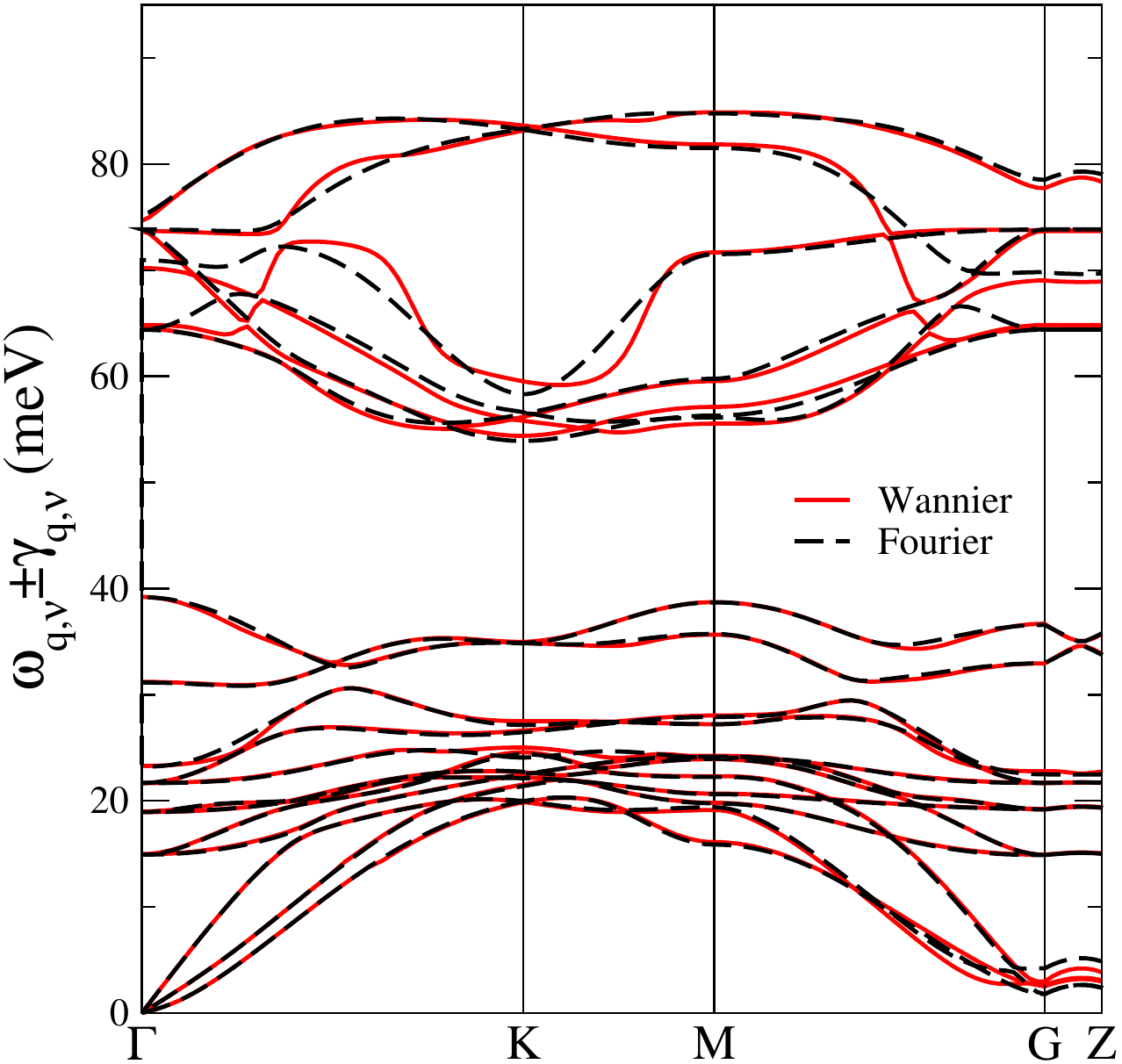}
\caption{
(Left): 
Phonon frequency (black line) and line-width (vertical red bars) for a perfect parabolic 2D electron gas with constant electron-phonon matrix element.
The phonon wavevector $q$ is measured either with respect to
zone center or with respect to ${\bf K}$.
(Right) : comparison between Wannier and Fourier interpolated
phonon dispersion in Li$_{1/18}$ZrNCl. 
}
\label{figs1}
\end{figure}

\begin{figure}[t]
\includegraphics[width=0.75\textwidth]{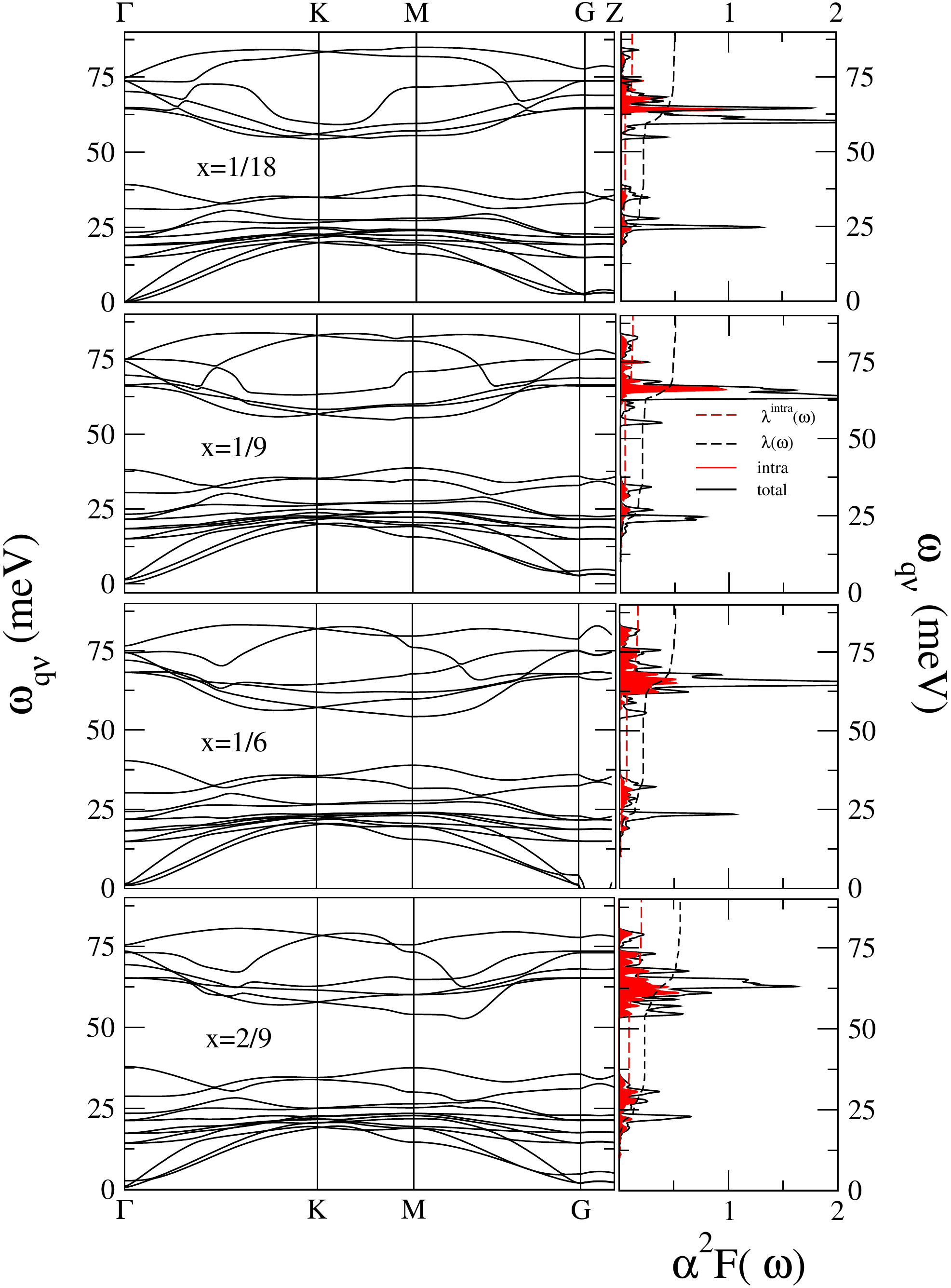}
\caption{Wannier interpolated phonon dispersion of  Li$_{x}$ZrNCl
as a function of doping. The total Eliashberg function and the
Eliashberg function due to intravalley scattering are also plotted
on the right panels.}
\label{figs2}
\end{figure}

In the case of constant matrix elements, the  phonon softening 
is ruled by the bare response-function $\chi_{0}({\bf q})$, see Eq. \ref{softconstant}.
Similarly the phonon linewidth is proportional to the 
nesting factor $N_f({\bf q})=\frac{1}{N_k}\sum_k \delta (\epsilon_{\bf k}-\epsilon_F)
\delta(\epsilon_{{\bf k}+{\bf q}}-\epsilon_F)$. 
In Fig.\ref{figs1} (left) we plot $\omega_{\bf K}+\widetilde{\Delta\omega}_{\bf K}$ 
and
the nesting factor (as vertical red bars) for a perfect parabolic 2D electron-gas.
In the right panel, we compare these results with the DFT calculations
for $x=1/18$. 
At this doping, as well for $x=1/9$ (see Fig.~\ref{figs2}) the $2k_F$
area around $\bf K$ 
is well separated from that around $\Gamma$.

In the ideal case of constant intervalley matrix elements, 
the phonon softening is flat in a region of radius $2k_F$ around  ${\bf K}$. 
Standard linear response calculations based on Fourier interpolation
do not reproduce this fact (see Fig.  \ref{figs1} on the right, black
dashed line) as the grid used in the calculation is too coarse.
On the contrary the analytical  behavior is very nicely captured by
our Wannier interpolation scheme \cite{CalandraWannier}, as shown
in Fig.  \ref{figs1} on the right, red lines.
The simple model also accounts for the behavior of the phonon linewidth. In this case the 
nesting factor as a function of ${\bf q}$ of Fig.\ref{figs1} (left)
should be compared with the phonon linewidth (red bars in Fig. 3 of the main paper).

The phonon dispersion and the Eliashberg functions for
several doping are shown in Fig. \ref{figs2}. The intravalley
contribution to the Eliashberg function and its integrated value
are plotted on the left panels. As it can be seen the Eliashberg
function is composed of two main peaks. The intravalley contribution
to the electron-phonon coupling is suppressed for small doping.

\begin{thebibliography}{99}
\bibitem{TecDet} The results reported in the present paper were obtained from
first-principles DFT in the linear response as
implemented in the QUANTUM-ESPRESSO~\cite{QE} package.
Converged phonon dispersion and electron-phonon interaction were
obtained by using Wannier interpolation  \cite{CalandraWannier}.
Li intercalation was simulated by changing the number of electrons and
adding a jellium compensating background~\cite{Heid}.
All the structures considered were obtained by using the experimental
lattice parameters and performing optimization of internal
coordinates. More details in \cite{SI}.
\bibitem{mustar}
The superconducting critical temperature was estimated using the
Allen-Dynes formula\cite{AllenDynes}.
We choose the value of  the unscreened Coulomb 
pseudopotential\cite{Bogolubov,Morel} ($\mu=0.198$) that leads to the correct estimate
of T$_c$ at the highest doping $x=2/9$ by using the RPA renormalized 
electron-phonon coupling ${\tilde \lambda}$. 
This value is also close
to the GW estimate of 0.203 computed at x=0.1 in Ref. \cite{Arita}
(page 13, middle second column).
We then obtain $\mu^*$ at any doping from 
$\mu^*=\mu/(1+\mu\log(\epsilon_f/\omega_D))$,
where $\epsilon_f$ and $\omega_D=90$ meV are the Fermi level and the Debye
energy respectively.
\end{thebibliography}
\end{document}